\documentclass[structabstract]{aa}  
\usepackage{graphicx}
\usepackage{txfonts}
\begin{document}
   \title{Molecules in the Circumstellar Disk Orbiting BP Piscium \\
     {\it Research Note}}


   \author{Joel H. Kastner
          \inst{1,2}
          \and
          B. Zuckerman\inst{3,4}
          \and
          Thierry Forveille\inst{1}
          }

   \institute{Laboratoire
  d'Astrophysique de Grenoble, Universit\'e Joseph Fourier --- CNRS,
  BP 53, 38041 Grenoble Cedex, France\\
              \email{joel.kastner@obs.ujf-grenoble.fr}
         \and
         Chester F. Carlson Center for Imaging
  Science, Rochester Institute of Technology, 54 Lomb Memorial Dr.,
  Rochester, NY 14623 USA
         \and
             Dept.\ of Physics \& Astronomy, University of California, Los
  Angeles 90095 USA 
         \and 
             UCLA Center for Astrobiology, University of California, Los
  Angeles 90095 USA
             }

\titlerunning{Molecules in the BP Psc Disk}
\authorrunning{Kastner, Zuckerman, \& Forveille}
   \date{Received ...; accepted ...}

 
  \abstract
  {BP Psc is a puzzling late-type, emission-line field star with large
    infrared excess. The star
    is encircled and enshrouded by a nearly edge-on, dusty
    circumstellar disk, and displays an extensive
    jet system similar to those associated with pre-main
    sequence (pre-MS) stars. However, the photospheric absorption features of
    the star itself appear more consistent with post-main sequence
    status.}
  {We seek to characterize the molecular gas component of the BP Psc
    disk, so as to compare the properties of its molecular disk with
    those of well-studied pre-main sequence stars.}
  {We conducted a mm-wave molecular line survey of BP Psc with the 30
    m telescope of the Institut de Radio Astronomie Millimetrique
    (IRAM). We use these data to investigate the kinematics, gas mass,
    and chemical constituents of the BP Psc disk.}
   {We detected lines of $^{12}$CO
  and $^{13}$CO and, possibly, very weak emission from HCO$^+$ and CN;
  HCN, H$_2$CO, and SiO are not detected. The CO line profiles of BP
  Psc are well fit by a model invoking a disk in Keplerian
  rotation. The mimumum disk gas mass, inferred from the $^{12}$CO
  line intensity and $^{13}$CO/$^{12}$CO line ratio, is $\sim$0.1
  Jupiter masses.  }
{The weakness of HCO$^+$ and CN (relative to $^{13}$CO) stands in
  sharp contrast to the strong HCO$^+$ and CN emission that
  characterizes most low-mass, pre-main sequence
  stars that have been the subjects of molecular emission-line
  surveys, and is suggestive of a very low level of X-ray-induced
  molecular ionization within the BP Psc disk. These results lend some
  support to the notion that BP Psc is an evolved star whose
  circumstellar disk has its origins in a catastrophic interaction
  with a close companion.}


   \maketitle
%

\section{Introduction}

Circumstellar disks around young stars serve both as the sources of
material for accreting young stars and as the sites of nascient
planets orbiting such stars.  Circumstellar disks around main sequence
and evolved stars may represent debris resulting from planetary
collisions, the destruction of planetary-mass companions, or similarly
devastating interactions with stellar-mass companions. 
Studies of such disks typically rely on dust emission (i.e., infrared
excess) to ascertain fundamental disk properties (e.g., disk
dimensions and mass; e.g., Backman \& Paresce 1993; Lagrange et al.\
2000; Zuckerman 2001). 
Complementary, sensitive
measurements of (sub)millimeter emission from CO (after H$_2$, the
second-most abundant molecular species) as well as less abundant
molecular species (such as HCN, CN, and HCO$^+$) toward
disk-enshrouded stars provide the best available means to determine
the residual molecular gas content and chemistry of circumstellar disks
(e.g., Zuckerman, Forveille \& Kastner 1995; Kastner et al.\ 1997;
Dutrey et al 1997; Thi et al 2004). Such measurements provide unique
tests for theories describing disk chemical and thermal evolution,
particularly for those models concerned with the effects of irradiation of
protoplanetary disks by high-energy photons -- UV and (perhaps more
importantly) X-rays -- on circumstellar disk chemistry and
energetics (e.g., Glassgold et al.\ 2004, 2007).

The enigmatic H$\alpha$ emission-line field star BP Psc (= StH$\alpha$
202; Stephenson 1986) stands as an
important object in this regard (Zuckerman et al.\
2008, hereafter Z08). Z08 established that this little-studied system
consists of a late G-type or early K-type star surrounded by a compact, dusty,
gaseous disk, and that the star-disk system is the source of highly
collimated jets. Z08 furthermore infer, on the basis of both its
classical, double-peaked CO line profiles and the dark-lane morphology
of near-IR adaptive optics images, that the BP Psc disk is viewed
nearly edge-on ($i\sim75^\circ$) and that the disk absorbs and
reradiates 75\% of the incident stellar luminosity as seen from Earth.

Like the intensively-studied, nearby classical T Tauri star (cTTS)
TW~Hya (Kastner et al.\ 1997; Webb et al.\ 1999; and references
therein), BP Psc is isolated and found at high galactic
latitude. Z08 show that if BP Psc is an early K-type pre-main
sequence (pre-MS) star then, like TW~Hya, it would be one of the closest (and
perhaps oldest) classical T Tauri stars known. However, unlike TW~Hya,
no young stellar association has been identified in the vicinity of BP
Psc (for a list of nearby [$D\stackrel{<}{\sim}100$ pc ] young
associations, see review in Zuckerman \& Song 2004). Indeed, Z08 also present
two lines of evidence suggesting that BP Psc may be an evolved star
--- most likely, a first-ascent giant: its $\lambda$6709.6 Li
absorption line is far weaker than in early K-type stars of age $<100$
Myr, and gravity-sensitive lines in its optical spectrum suggest it is
of luminosity class IV or III, i.e., its surface gravity is lower than
a typical 10 Myr-old, K-type, pre-MS star. Whether BP Psc is an
isolated, relatively old cTTS or a post-MS star undergoing an
episode of collimated mass loss, its nearly edge-on disk and its
system of jets and Herbig-Haro objects --- which is as spectacular as
the jet systems commonly associated with very young (still
cloud-embedded) pre-MS stars --- make BP Psc an exceedingly unusual
object.

To better characterize the disk orbiting BP Psc, we undertook a
mm-wave molecular line survey with the 30 m telescope of the Institut
de Radio Astronomie Millimetrique (IRAM\footnote{http://iram.fr/}). In
addition to constraining the mass, kinematics, and chemistry of the BP
Psc disk, the results point out significant differences between this
system and those pre-MS star molecular disks that have also been the
subjects of extensive radio emission-line surveys.

\section{Observations}

\subsection{Data Acquisition and Reduction}

\begin{table*}
\begin{center}
\caption{Molecular emission lines measured toward BP Psc}
\renewcommand{\footnoterule}{}  
\begin{tabular}{ccccccc}
\hline \hline
Transition & $\nu$ & $T_{B,max}$ & $I$ & $v_d$ & $q$  & $p_d$ \\
 & (GHz) & (mK) & (K km s$^{-1}$) & (km s$^{-1}$) & & \\
 (1) & (2) & (3) & (4) & (5) & (6) & (7) \\
\hline
$^{12}$CO (1--0) & 115.270 & 85 (4) & 1.2 (0.1) & 5.3 (0.3) & 0.8 (0.1) & 0.33: \\  
$^{12}$CO (2--1) & 230.538 & 260 (35) & 3.5 (0.4) & 4.7 (0.3) &  0.7 (0.1) & 0.1: \\ 
$^{12}$CO (3--2) & 345.796 & 530 (120) & 5.2 (0.8) & 4.0 (0.5) &  0.5:  & 0.1: \\ 
$^{13}$CO (2--1) & 220.399 & 33 (2) & 0.4 (0.1) & 4.0 (0.9) & 0.9 (0.2) & ... \\  
HCO$^+$ (3--2) & 267.558 & $<15$ & $<0.23$ & ... & ... & ... \\  
CN (2--1) & 226.875 & $<9$ & $<0.15$ & ... & ... & ... \\
SiO (v=1, 2--1) & 86.243 & $<12$ & $<0.18$ & ... & ... & ... \\  
HCN (1--0) & 88.632 & $<$4 & $<0.06$ & ... & ... & ... \\  
HCN (3--2) & 265.886 & $<$70 & $<1.1$ & ... & ... & ... \\
H$_2$CO (2$_{21}$--1$_{11}$) & 140.839 & $<$10  & $<0.15$ & ... & ... & ... \\
\hline
\end{tabular}
\end{center}

\footnotesize{NOTES --- Data obtained in 2007 December with the IRAM 30 m,
  with exception of $^{12}$CO (2--1) (obtained 1996 May with the IRAM 30
  m) and $^{12}$CO (3--2) (obtained 1996 Feb.\ with the JCMT 15
  m). Peak line intensity ($T_{B,max}$), integrated line
  intensity ($I$), one-half peak-to-peak velocity difference ($v_d$),
  radial temperature power law index ($q$), and outer disk cutoff
  parameter ($p_d$) obtained from fits to disk model line
  profiles. Numbers in parentheses indicate formal (1$\sigma$)
  uncertainties in best-fit parameter values. Upper limits on
  $T_{B,max}$ and $I$ listed for nondetected transitions of SiO, HCN,
  and H$_2$CO are 3$\sigma$, based on measured channel-to-channel
  noise levels in off-line regions of spectrum and an assumed
  linewidth of 15 km s$^{-1}$. See \S\S 2.2 and 3.1.}

\end{table*}


We conducted our molecular line survey of BP Psc with the IRAM 30 m
telescope during the period 4--6 Dec.\ 2007. The molecules and
transitions observed are listed in Table 1. We observed simultaneously
in either the 100 GHz (3 mm) and 230 GHz (1 mm) or 150 GHz (2 mm) and
270 GHz (1 mm) bands, and in both polarizations in each band, using
receiver combinations A100+B100 and C230+D230 or A150+B150 and
C270+D270 (all in SSB mode), respectively. The 1 MHz filter banks
served as the spectral line backends. The weather was excellent to good
($\tau_{225} \sim 0.1$ to 0.3) throughout the period; time-averaged
system temperatures in both the 3 mm and 1 mm bands were in the range
250--400 K. We checked pointing and focus (using Uranus as the
reference) every 1-2 hours, and both were found to be stable and
reliable; typical pointing errors were $<3''$, i.e., less than 1/8
beamwidth for 3 mm (FWHP beamwidth $21''$) and less than 1/4 beamwidth
for 1 mm (FWHP beamwidth $12''$). Individual spectral scans were of
duration 200 s, with total integration times (per polarization)
ranging from $\sim40$ minutes (for HCN(3--2)) to $\sim8.5$ hours (for
HCO$^+$(3--2)).

We used the CLASS\footnote{See http://iram.fr/IRAMFR/GILDAS/} radio
spectral line data 
reduction package to sum all individual spectral scans obtained in
both polarizations for a given transition, and then to subtract a
linear-fit baseline from each of these integrated spectra, calculating
channel-to-channel noise levels in the process. A few
individual scans were discarded due to baseline anomalies. 
All antenna temperature measurements reported in Table 1 (i.e.,
$T_{B,max}$ and $I$; see \S 3.1), have been corrected for beam
efficiency assuming $B_{eff} = 0.76, 0.70, 0.57$ and
0.45 at observing frequencies of 86, 115, 230, and 270 GHz,
respectively\footnote{See http://iram.fr/IRAMFR/ARN/aug05/node6.html}.

\subsection{Results}

\begin{figure}
\includegraphics[width=9cm,angle=0]{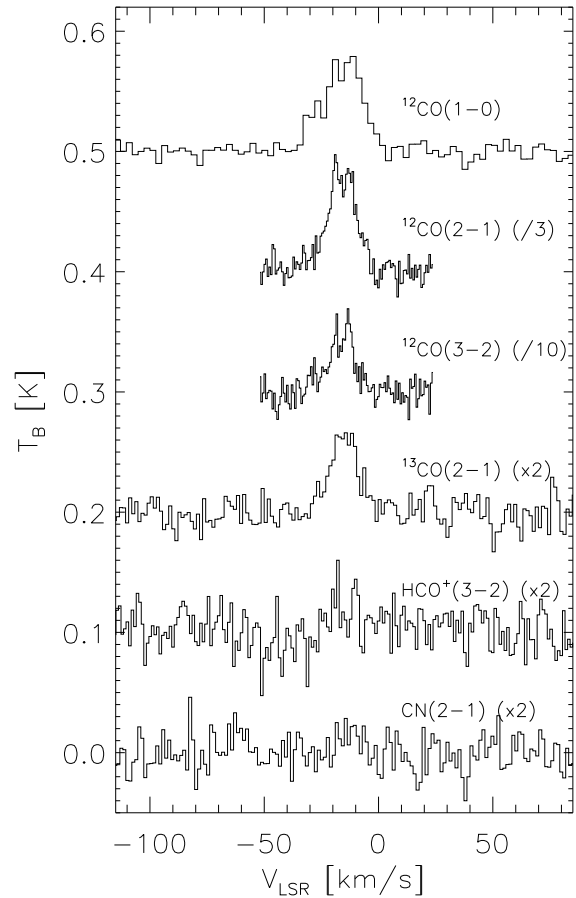}
\caption{Radio (mm-wave) molecular spectra of BP Psc. Ordinate is
  velocity with respect to the 
  Local Standard of Rest (LSR) and abscissa is antenna temperature
  corrected for beam efficiency. Spectral baselines are offset 
  in T$_B$ for clarity. All spectra except $^{12}$CO(3--2) were
  obtained with the IRAM 30 m; $^{12}$CO(3--2) was obtained at the
  JCMT 15 m. See text and Table 1.  }
\label{fig:specFig}
\end{figure}

Results are summarized in Table 1 and Fig.\ 1. Including the previous
observations obtained with the IRAM 30 m and the 15 m James Clerk
Maxwell Telescope (JCMT) reported in Z08, the BP Psc molecular line
survey has yielded detections of $^{12}$CO(1--0), $^{12}$CO(2--1),
$^{12}$CO(3--2) and $^{13}$CO(2--1) emission; only tentative
detections of HCO$^+$(3--2) and CN(2--1) emission; and nondetections
of HCN, H$_2$CO, and SiO (maser) emission. Given the marginal
($\sim$2$\sigma$) significance of the possible detections of the
HCO$^+$(3--2) and CN(2--1) lines, these observations are reported as
upper limits in Table 1 and are considered as such in the discussion
below. For HCN, H$_2$CO, and SiO, the upper limits on peak antenna
temperature $T_{B,max}$ and integrated line intensity $I$ listed in
Table 1 were obtained from the channel-to-channel noise level measured
via spectral baseline fitting, assuming a linewidth of 15 km s$^{-1}$
FWHM (as estimated from CO line profile fitting; \S 3.1). The
measurements of the CO line profile parameters and of the upper limits
on $T_{B,max}$ and $I$ for the marginal detections of HCO$^+$ and CN
are described in \S 3.1.  Relative to the intensities of the CO lines,
the HCO$^+$ and CN emission from the BP Psc disk is evidently quite
weak in comparison to pre-MS star disks (Dutrey et al.\ 1997; Thi et
al.\ 2004; see \S 3.3). The upper limits on HCN emission in
Table 1 are not similarly constraining.

\section{Analysis and Discussion}

\subsection{CO line profiles: constraints on disk structure and
  kinematics}

Some of the CO lines displayed in Fig.\ 1 (see also Z08)
appear to display the double-peaked profile characteristic of a
circumstellar disk in Keplerian rotation (e.g., Beckwith \& Sargent
1993, hereafter BS93; Omodaka et al.\ 1992). In contrast, the
$^{13}$CO(2--1) line appears flat-topped, and is reminiscent of the
$^{12}$CO(3--2) line profile measured with the SubMillimeter Array
(Z08). All of the CO lines display broad wings extending $\pm\sim15$
km s$^{-1}$ to either side of the systemic velocity of BP Psc
($V_{LSR} = -15$ km s$^{-1}$). In addition, the $^{12}$CO(1--0) line
appears to display a broad negative-velocity shoulder; a feature
near $V_{LSR} \sim -25$ km s$^{-1}$ may be present in the other CO
profiles as well. 

The CO line profile asymmetries, and the slight variation from transition to
transition, suggests departures from the ``ideal'' flattened disk with
sharp outer edge in Keplerian rotation. Nevertheless, the BP Psc CO
profiles can be reasonably well described by a simple parameterization
of the results of detailed numerical models of Keplerian molecular
disks surrounding T Tauri stars (Eqs.\ 27, 28 of BS93). The
parameterization applies in the case of a disk viewed edge-on --- a
reasonable approximation of the BP Psc viewing geometry --- and is in
terms of the Keplerian velocity $v_d$ at the outer edge of the disk
($v_d = (GM_\star/R_d)^{1/2}$, where $M_\star$ is the stellar mass and
$R_d$ the disk outer radius) and the index $q$ of the (assumed) power
law dependence of disk gas temperature $T$ on radial position $r$
(i.e., $T(r) \propto r^{-q}$). Hence, $v_d$ defines the peaks of the
``twin horns'' in the ideal line profile, with the value of $v_d$
roughly equal to half the peak-to-peak velocity separation of the
horns. Because the velocity dependence of line flux is $F(v) \propto
v^{3q-5}$ for $v>v_d$ (BS93), $q$ dictates the steepness of the line
wings in the model profiles. Values of $q$ larger than the canonical
$q=1/2$ yield an ``excess'' of high-$v$ disk material relative to the
``standard'' BS93 disk models and, hence, broader wings than those of
the BS93 profiles. In modeling the BP Psc CO line profiles, we
modified the low-$v$ BS93 profile parameterization (BS93, Eq.\ 28) by
introducing a variable index for the power-law dependence of the line
intensity on velocity (i.e., $F(v) \propto v^{p_d}$ for $v<v_d$).
This parameter, $p_d$, in effect accounts for the fact that the disk
likely does not have a sharp cutoff at $R_d$; values of $p_d<1.0$ tend
to fill in the central regions of the line profile.

Fitting the BP Psc CO lines with this simple model thereby allows the
empirical determination of $v_d$ and the peak line intensity
$T_{B,max}$, as well as the temperature profile and outer edge cutoff
power-law indices $q$ and $p_d$. For the three $^{12}$CO lines, all
four parameters were left free during the model fitting. For the
(lower signal-to-noise) $^{13}$CO(2--1) line, the value of $p_d$ was
fixed to the value estimated from the $^{12}$CO(2--1) profile fitting
(see below). The total CO line intensities ($I$) were then obtained by
integrating the best-fit models over the velocity range $-15$ to $+15$
km s$^{-1}$ with respect to the systemic velocity of BP Psc. For the
(marginal significance) CN and HCO$^+$ lines, all parameters except
$T_{B,max}$ were fixed to the values obtained from the fit to the
$^{12}$CO(2--1) profile, such that $T_{B,max}$ was left as the only
free parameter; the 3$\sigma$ upper limits on $T_{B,max}$ and $I$ in
Table 1 are then based on the formal uncertainties in the resulting fit
to $T_{B,max}$ (these upper limits are similar to those
estimated from the baseline fitting procedure).

\begin{figure*}
\includegraphics[scale=0.45,angle=0]{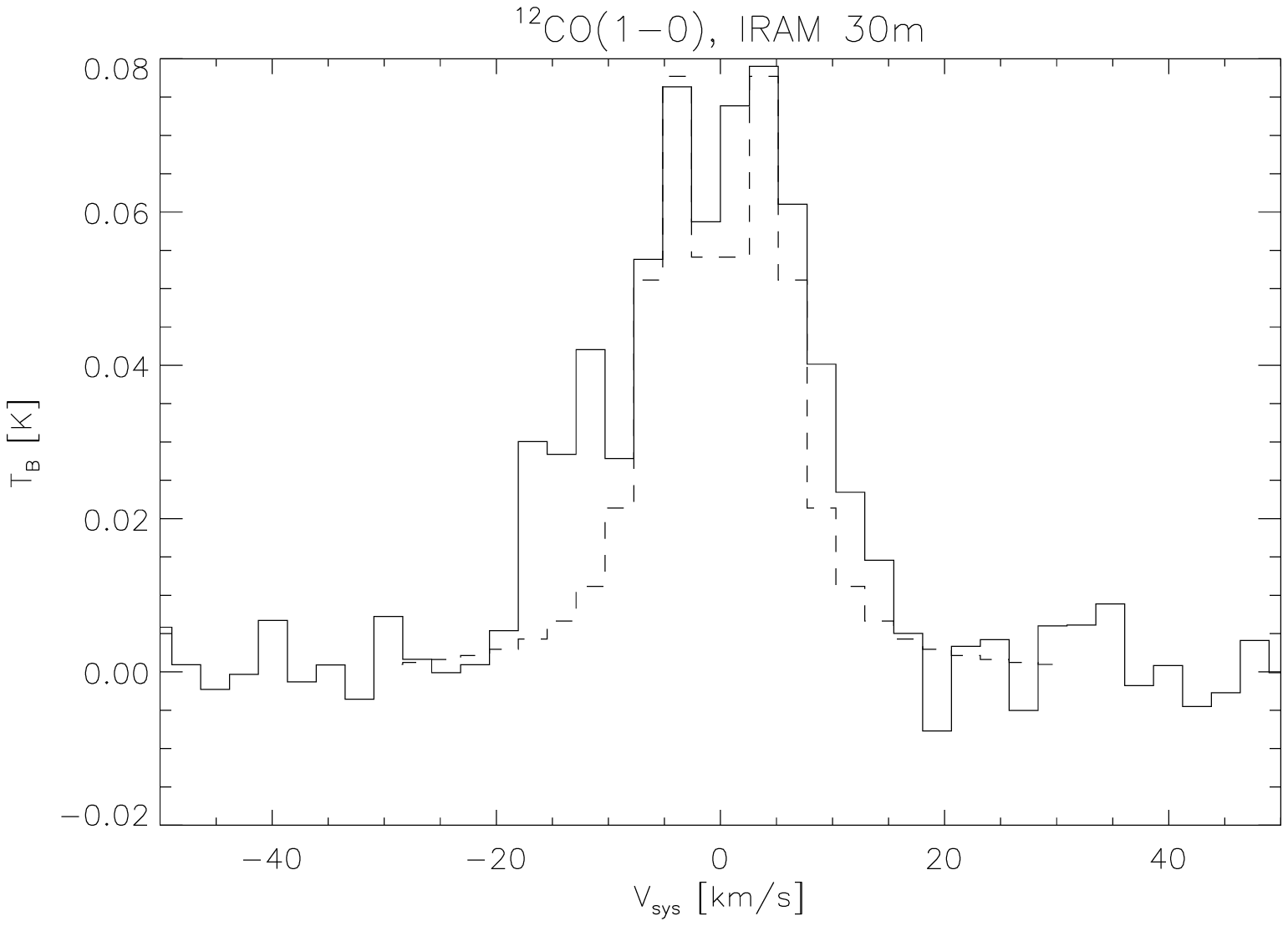}
\includegraphics[scale=0.45,angle=0]{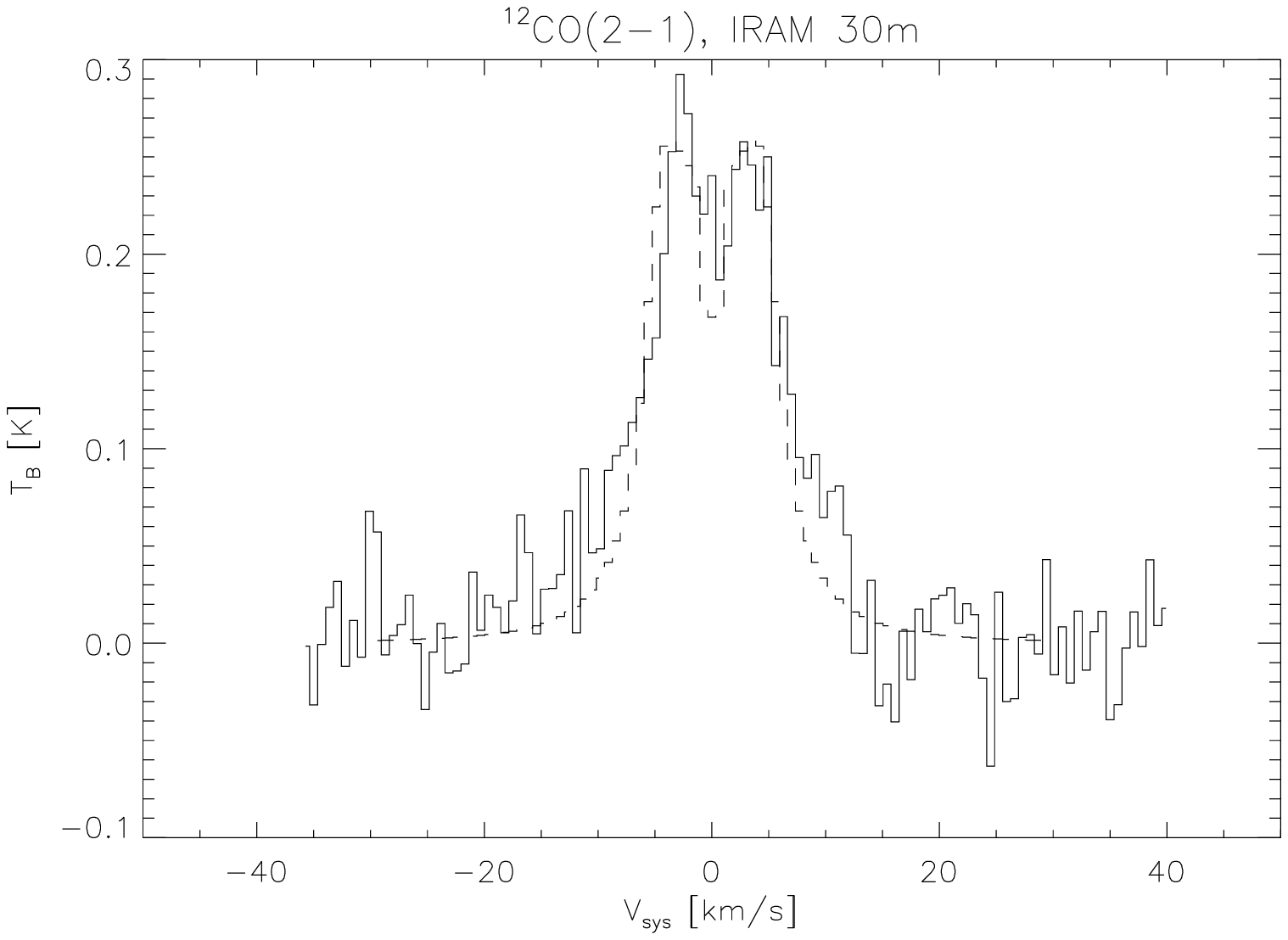}
\includegraphics[scale=0.45,angle=0]{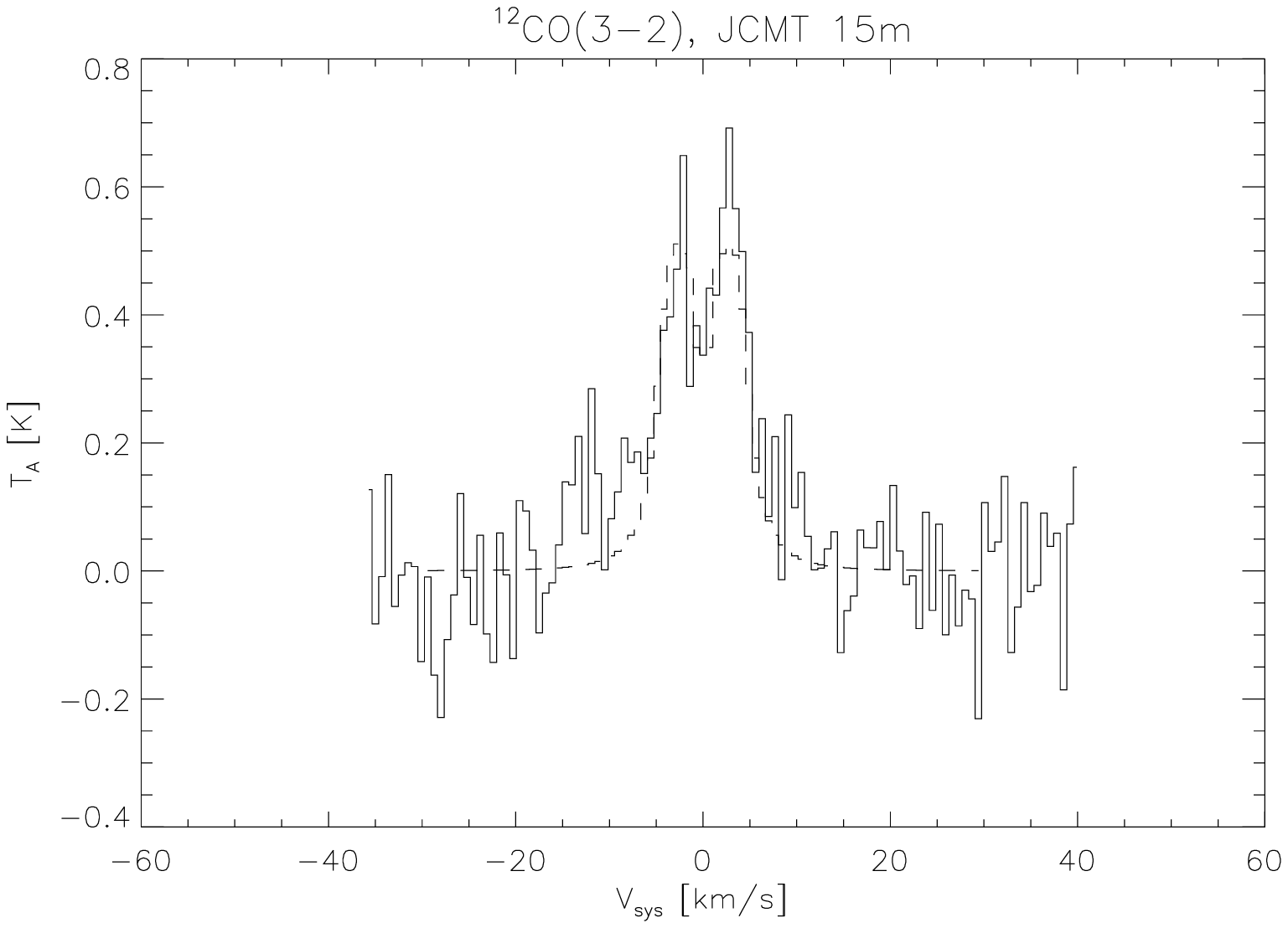}
\includegraphics[scale=0.45,angle=0]{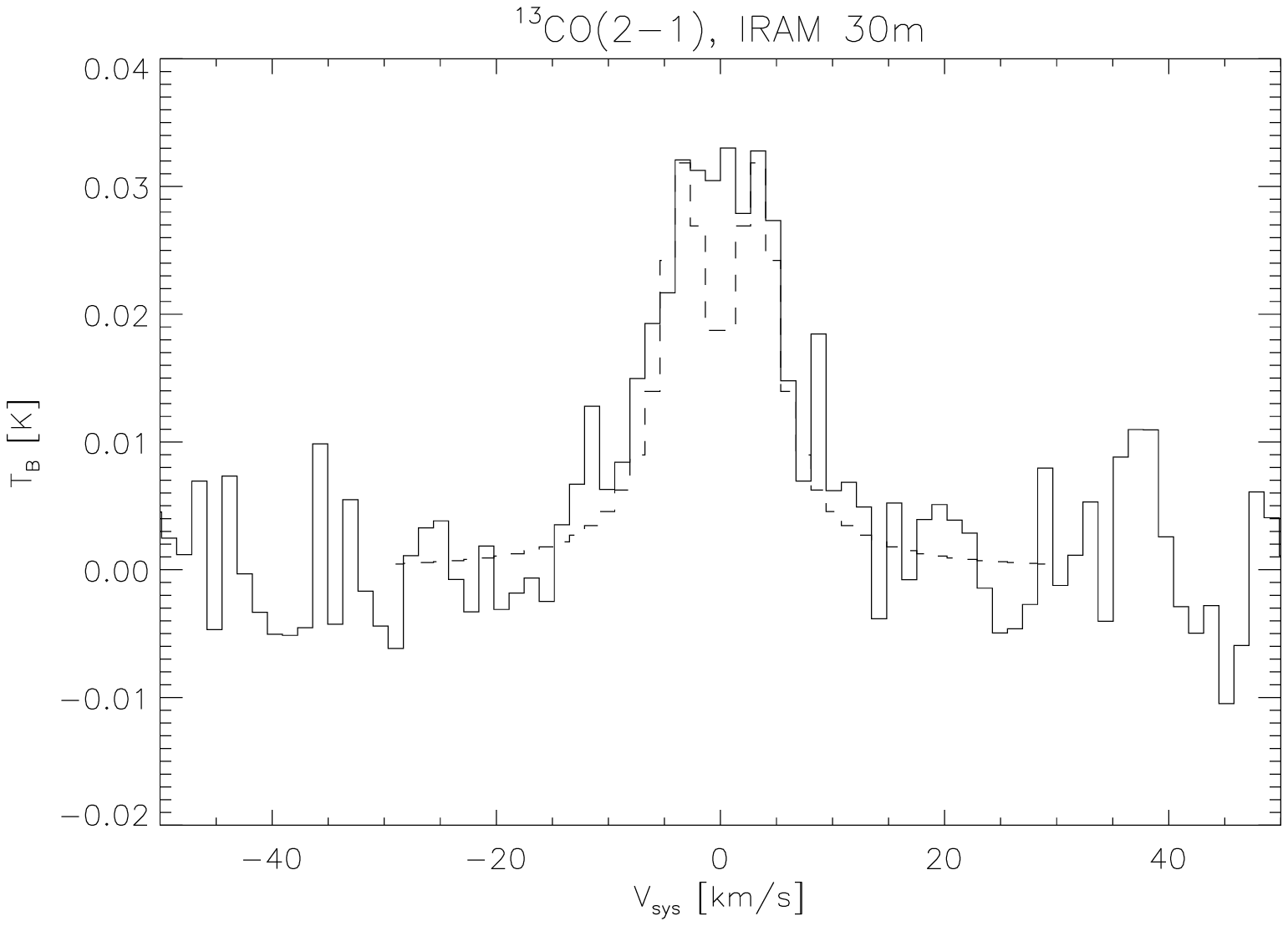}
\caption{CO line profiles of BP Psc (solid lines) overlaid with
  best-fit Keplerian disk model profiles (dashed lines). Ordinate is
  velocity with respect to the systemic velocity of BP Psc 
  ($V_{LSR}=-15$ km s$^{-1}$).}
\label{fig:COprofiles}
\end{figure*}

The results of this profile fitting exercise are listed in cols.\ 3--7
of Table 1. In Fig.\ 2 we display the best-fit model CO profiles
overlaid on the IRAM and JCMT spectra. For three of the four CO lines
measured thus far, we obtain best-fit values of $v_d$ in the range
4.0--4.7 km s$^{-1}$. The $^{12}$CO(1--0) line yields a somewhat
larger best-fit value of $v_d = 5.3\pm0.3$ km s$^{-1}$. These values
of $v_d$ are all systematically larger than those obtained from
simple, two-Gaussian model fits to the single-dish (Fig.\ 1) and
interferometer (Z08) CO profiles, which yield peak-to-peak separations
in the range 6.0--7.5 km s$^{-1}$ (i.e., $v_d$ in the range 3.0--3.75
km s$^{-1}$). The discrepancy is due to the fact that, in the BS93
model parameterization, the value of $v_d$ corresponds to the outer
edges of the ``twin peaks'' in the line profile, whereas the Gaussian
fits find the velocity centroids of these peaks. 

We (tentatively) conclude that $v_d$ lies in the range 3.0--4.0 km
s$^{-1}$. For such a range of $v_d$, the mass of the central star
$M_\star$ would lie in the range 0.5--0.9 (1.0--1.8) $M_\odot$, given
a distance of 100 (300) pc (Z08). The Keplerian profile fits also indicate
values of $q$ and $p_d$ in the ranges 0.7--0.9 and 0.1-0.3,
respectively (Table 1), suggesting that the gas temperature in the BP
Psc disk falls off somewhat more steeply than $r^{-1/2}$ and that the
disk does not have a sharp outer edge. 

\subsection{Disk gas mass, density, and gas-to-dust ratio}

Due to lack of knowledge of the optical depth in $^{12}$CO(2--1), Z08
were only able to estimate an approximate lower limit to the mass of
gas in the BP Psc disk. Detection of $^{13}$CO(2--1) emission (Fig.\
1; Table 1), in combination with our previous measurement of
$^{12}$CO(2--1) at the 30 m (Z08), allows us to refine this disk gas
mass estimate.  Specifically, the $^{12}$CO(2--1)/$^{13}$CO(2--1) line
ratio is now determined to be 9$\pm$2 (Table 1). Assuming the solar
value of 89 for the number ratio of $^{12}$C:$^{13}$C, and that the
$^{13}$CO(2--1) emission is optically thin (see below), this
implies an optical depth in the $^{12}$CO(2--1) line of
$\tau\sim10$. With this $\tau$ estimate, and adopting standard
formalism (e.g., Eq.\ 4 in Z08) and standard assumptions (i.e., a mean
gas temperature of $\sim40$ K and a CO:H$_2$ number ratio of
10$^{-4}$), we estimate a disk gas mass of $\sim10^{-4}$ $M_\odot$
($\sim0.1$ Jupiter masses) assuming that BP Psc is a pre-MS star at a
distance of $\sim$100 pc. The disk would contain about a Jupiter mass
of gas if BP Psc were instead a first-ascent giant at 300 pc.

Adopting the Z08 estimate for the mass of cold ($T<200$ K) dust in the
disk, 0.7 Earth masses (for an assumed distance of 100 pc), these gas
mass estimates imply a (distance-independent) gas-to-dust ratio of
$\sim50$. Meanwhile, assuming BP Psc is pre-MS, such that its disk
outer radius is $R_d = 50$ AU and disk scale height is 5 AU (the
latter based on the heavy obscuration of the central star for an
assumed inclination $\sim75^\circ$; Z08), the implied mean H$_2$
number density is $n_{H_2}\sim2\times10^8$ cm$^{-3}$. This mean
density --- which depends only weakly on the assumed distance to,
hence evolutionary status of, BP Psc --- is significantly larger than
the critical densities of excitation of all of the molecular
transitions in Table 1 (see, e.g., Table 1 of Dutrey et al.\ 1997).

Our disk gas mass, gas-to-dust ratio, and mean H$_2$ number density
estimates remain uncertain --- and may still only represent lower
limits --- given the possibility that most CO molecules in the disk
are frozen out onto (i.e., trapped in icy mantles surrounding) dust
grains, or are preferentially photodissociated relative to
H$_2$. Indeed, the assumptions invoked above for the BP Psc disk --- optically
thin $^{13}$CO emission and CO:H$_2$ number ratio of 10$^{-4}$ --- may
be mutually exclusive. Adopting the standard ISM gas-to-dust ratio of 100, Thi
et al. (2004) and Dutrey et al. (1997) estimate that, in pre-MS disks,
CO gas is depleted by factors $\sim$10--200. Thi et al. (2004) further
estimate $^{13}$CO optical depths $< 1$ for each of the 4 pre-MS disks
that they observed with the JCMT, whereas the BS93 CO line profile
models indicate that the $^{13}$CO optical depth should be $>> 1$
given a gas-to-dust mass ratio of 100 and CO:H$_2$ $= 10^{-4}$ by
number. 


\subsection{Weakness of HCO$^+$ and CN: inefficient molecular
  ionization?}

Models indicate that both HCO$^+$ and CN should be sensitive tracers
of X-ray molecular ionization rate at a given gas column density.
Glassgold et al. (2004) note that the abundance of HCO$^+$ is likely
to be sharply elevated in pre-MS circumstellar disks that are
irradiated by X-rays from the vicinity of the central star. Although
the Glassgold et al.\ disk models do not include HCN or CN, the
abundance of the latter (the dissociation product of HCN) should also
be enhanced by exposure to stellar X-rays (Lepp \& Dalgarno
1996). Kastner et al.\ (1997) and Thi et al (2004) considered both
X-rays and UV as potential drivers of large HCO$^+$/CO and CN/HCN
abundance ratios (relative to the molecular cloud values of these
ratios) measured in T Tauri and Ae/Be star disks.  Ionization of H$_2$
by soft ($\sim1$ keV) X-rays was also implicated as the source of
enhanced HCO$^+$ in the molecular envelope surrounding the planetary
nebula NGC 7027 (Deguchi et al.\ 1990), an interpretation supported by
the subsequent detection of an extended region of luminous, soft X-ray
emission within this object (Kastner et al.\ 2001).

\begin{figure}
\includegraphics[width=6cm,angle=90]{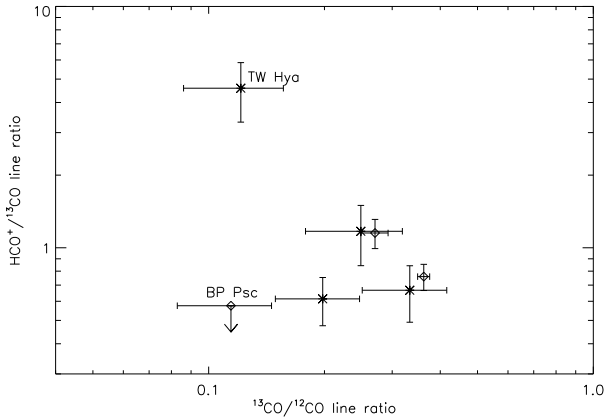}
\caption{HCO$^+$/$^{13}$CO vs.\ $^{13}$CO/$^{12}$CO line ratios for BP
  Psc and for the (6) pre-MS stars measured to date (including the
  evolved cTTS TW Hya). Data points corresponding to JCMT 15 m
  measurements of the integrated line intensities in the HCO$^+$(4-3)
  and $^{13}$CO(3--2) transitions (Thi et al.\ 2004) are indicated with
  asterisks; points corresponding to IRAM 30 m measurements of the
  integrated line intensities in the HCO$^+$(3-2) and $^{13}$CO(2--1)
  transitions are indicated with diamonds (BP Psc, this paper; other
  stars, Dutrey et al.\ (1997).}
\label{fig:CO_HCO+_Ratios}
\end{figure}

\begin{figure}
\includegraphics[width=6cm,angle=90]{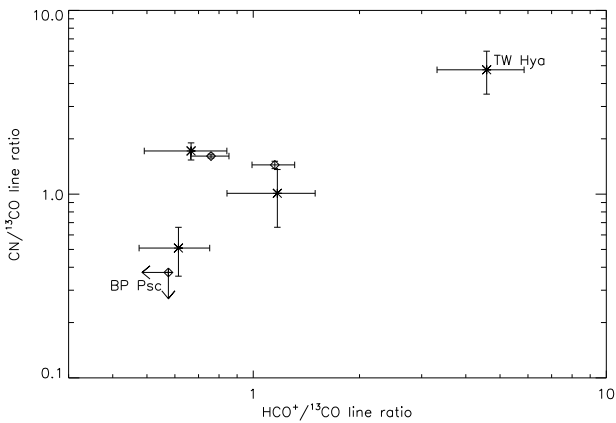}
\caption{CN/$^{13}$CO vs.\ HCO$^+$/$^{13}$CO line ratios for BP Psc
  and pre-MS stars. Data points corresponding to JCMT 15 m
  measurements of the integrated line intensities in the HCO$^+$(4--3),
  CN(3--2) and $^{13}$CO(3--2) transitions (Thi et al.\ 2004) are indicated
  with asterisks; points corresponding to IRAM 30 m measurements of
  the HCO$^+$(3-2), CN(2--1), and $^{13}$CO(2--1) transitions are indicated with
  diamonds (BP Psc, this paper; other stars, Dutrey et al.\
  (1997). The lone data point near the BP Psc upper limits is the
  Herbig Ae star MWC 480.} 
\label{fig:HCO+_CN_ratios}
\end{figure}

In Figs.~\ref{fig:CO_HCO+_Ratios}, \ref{fig:HCO+_CN_ratios} we plot
the Table 1 results for $^{13}$CO/$^{12}$CO, HCO$^+$/$^{13}$CO, and
CN/$^{13}$CO line ratios measured for BP Psc along with the same
ratios for all (6) other circumstellar disk sources for which line
intensities have been published to date (TW Hya: Kastner et al.\ 1997
and Thi et al.\ 2004; DM Tau and GG Tau: Dutrey et al.\ 1997; LkCa 15,
HD 163296, and MWC 480: Thi et al.\ 2004). The ratios have been
calculated from line intensity data obtained at transitions that lie
within $\sim15$\% or less in frequency, so are relatively insensitive
to beam dilution effects; these and other systematic errors in the
line ratios (e.g., corrections for the different relative
  contributions from unmeasured CN hyperfine structure lines at each
  rotation transition; Skatrud et al.\ 1983) are similar to or smaller
than the typical measurement uncertainties. Assuming the $^{13}$CO
emission from all stars is optically thin, the $^{13}$CO/$^{12}$CO
line ratio should provide a measure of $^{12}$CO column density (\S
3.2). Meanwhile, because the typical disk gas densities are sufficient
to well-excite the observed transitions (\S 3.2), the
HCO$^+$/$^{13}$CO and CN/$^{13}$CO ratios serve as measures of the
relative abundances (as opposed to the ease of excitation) of HCO$^+$
and CN, respectively (again assuming $^{13}$CO is optically thin).

In both Fig.~\ref{fig:CO_HCO+_Ratios} and
Fig.~\ref{fig:HCO+_CN_ratios}, five of the six previously observed
pre-MS stars appear clustered together. The outlier among the pre-MS
stars is the ``old'' (age $\sim8$ Myr) cTTS TW Hya. Although $^{12}$CO
is optically thick in all of the objects, the $^{12}$CO optical depth
of the TW Hya disk is smaller than that of the other pre-MS stars (see
also Table 8 of Thi et al.\ 2004), likely reflecting its relatively
evolved state. TW Hya also displays the largest HCO$^+$/$^{13}$CO and
CN/$^{13}$CO line ratios among the pre-MS stars.

Fig.~\ref{fig:HCO+_CN_ratios} is indicative of a correlation between
the HCO$^+$/$^{13}$CO and CN/$^{13}$CO line ratios in pre-MS
circumstellar molecular disks. Furthermore, the star with the largest
HCO$^+$/$^{13}$CO and CN/$^{13}$CO line ratios, TW Hya, exhibits the
largest quiescent X-ray luminosity: $L_X = 1.4\times10^{30}$ erg
s$^{-1}$ (Kastner et al.\ 2002), compared with
$L_X\stackrel{<}{\sim}5\times10^{29}$ erg s$^{-1}$ for those stars for
which X-ray data have been published to date (HD 163296, Stelzer et
al.\ 2006; DM Tau, G\"{u}del et al.\ 2006; GG Tau, Stelzer \&
Neuhauser 2001). TW Hya also possesses the smallest molecular disk
radius and mass (Dutrey et al.\ 1997, Thi et al.\ 2004, and references
therein). Hence the apparent correlation of HCO$^+$/$^{13}$CO and
CN/$^{13}$CO, combined with the inferred large (2--3 order of
magnitude) enhancement of the CN/HCN and HCO$^+$/CO abundance ratios
in all of the pre-MS disks, relative to values of these ratios in
molecular cloud cores (Thi et al 2004), supports the interpretation
that high molecular ionization rates --- most likely due to
irradiation by X-rays emitted from stellar coronae and/or from
star-disk interfaces --- enhance the abundances of both HCO$^+$ and
CN in these disks.

If disk ionization by central X-ray sources is
responsible for the potential correlation between HCO$^+$/$^{13}$CO
and CN/$^{13}$CO apparent in Fig.~\ref{fig:HCO+_CN_ratios}, then this
correlation indicates that the molecular gas disks lie in a regime
where both HCO$^+$ and CN abundances are roughly proportional to X-ray
ionization rate (Lepp \& Dalgarno 1996, their Figs.~2, 3). More
specifically --- noting that the HCO$^+$/$^{12}$CO and CN/$^{12}$CO
number ratios in the pre-MS disks are inferred to be as large as
$\sim3\times10^{-4}$ and $\sim2\times10^{-3}$, respectively (Thi et
al.\ 2004) --- the Lepp \& Dalgarno models indicate that ionization
rates in the 6 previously measured pre-MS disks lie in the range
$10^{-15}$--$10^{-13}$ s$^{-1}$ (given a representative disk number
density $n\sim10^7$ cm$^{-3}$ for the molecular line-emitting regions;
Thi et al.\ 2004). This range is several orders of magnitude larger
than the canonical molecular ionization rate due to cosmic rays.

The $^{13}$CO/$^{12}$CO ratio of BP Psc is similar to that of TW Hya,
and is smaller than the ratios characteristic of the (younger) pre-MS
stars in Fig.~\ref{fig:CO_HCO+_Ratios}.  If BP Psc were pre-MS, this
comparison would suggest that the BP Psc disk is as highly evolved as
the disk orbiting TW Hya. However, the HCO$^+$/$^{13}$CO and
CN/$^{13}$CO line ratio upper limits measured for BP Psc are at least
a factor $\sim$30 lower than those of TW Hya and a factor $\sim$3--10
lower than all but one of the other pre-MS stars. The only
  pre-MS star near which BP Psc may lie in
  Fig~\ref{fig:HCO+_CN_ratios} is the (intermediate-mass) Herbig Ae
  star MWC 480 --- a star with which BP Psc, if a (K-type, low-mass)
  pre-MS star, otherwise would have little in common. If the trend
observed in Fig~\ref{fig:HCO+_CN_ratios} is indeed indicative of X-ray
ionization of H$_2$, then its low HCO$^+$/$^{13}$CO and CN/$^{13}$CO
ratios imply BP Psc has an anomalously low X-ray flux at its disk
surface, compared with the other (pre-MS) star-disk systems (with
  the possible exception of MWC 480).

Unfortunately, the only X-ray observation of BP Psc obtained thus far
--- a nondetection in the ROSAT All-Sky Survey (RASS) --- cannot be
used to test this hypothesis. The RASS nondetection (PSPC count rate
$\stackrel{<}{\sim} 0.1$ s$^{-1}$) implies an intrinsic X-ray flux
upper limit $F_X \stackrel{<}{\sim}9\times10^{-12}$ erg s$^{-1}$
cm$^{-2}$ (0.1--2.0 keV) assuming\footnote{The characteristic $T_X$ of
  BP Psc would be lower than $10^7$ K if its X-ray spectrum resembles
  that of TW Hya ($T_X\sim3\times10^6$ K; Kastner et al.\ 2002). If
  so, the RASS nondetection of BP Psc would provide even poorer
  constraints on its intrinsic X-ray flux $F_X$ and, hence, $L_X$.}
$T_X = 10^7$ K and an intervening absorbing column $N_H =
8\times10^{21}$ cm$^{-2}$ (adopting the $N_{H_2}$ value obtained by
Z08 and correcting for the $^{12}$CO optical depth determined in \S
3.1), or an X-ray luminosity $L_X \stackrel{<}{\sim}10^{31}$ erg
s$^{-1}$ for an assumed source distance of $\le100$ pc. Hence, if BP
Psc is a pre-MS star, the RASS nondetection would be consistent with a
quiescent $L_X$ even larger than that of TW Hya ($1.4\times10^{30}$
erg s$^{-1}$; Kastner et al.\ 2002). If BP Psc is, instead, a post-MS
G star at a distance $\sim300$ pc (Z08), the RASS nondetection does
not preclude the possibility that its $L_X$ is comparable to that of
the more X-ray-luminous G-type giants (Gondoin 2005).

\section{Conclusions}

The suite of molecular line data obtained thus far for BP Psc (Table
1; Fig.\ 1) confirms that its circumstellar disk in certain respects
resembles those of pre-MS stars --- but also reveals some fundamental
differences. The BP Psc CO line profiles indicate Keplerian rotation
of at least $\sim0.1$ Jupiter masses of disk gas around a central
star(s) whose mass lies the range 0.5--0.9 $M_\odot$, assuming
pre-main sequence status (\S\S 3.1, 3.2 and Z08), although the
profiles are also suggestive of a disk temperature gradient that is
somewhat steeper than the canonical $r^{-1/2}$ characteristic of
pre-MS star molecular disks (BS93). The $^{12}$CO optical depth of the
BP Psc disk (as inferred from its $^{13}$CO(2--1)/$^{12}$CO(2--1) line
ratio) is similar to that of the highly evolved (age $\sim8$ Myr) cTTS
TW Hya, consistent with the interpretation that --- like TW Hya --- BP
Psc is an isolated, ``old'' (yet actively accreting) TTS.

On the other hand, the HCO$^+$/$^{13}$CO and CN/$^{13}$CO line
ratios of BP Psc are smaller than
those of most (if not all) pre-MS star disks observed to date in
these molecules. In particular, its disk chemistry differs sharply
from that of TW Hya, its presumed closest pre-MS analog (Fig.\
4). Indeed, in this key respect, the circumstellar molecular disk of
BP Psc would appear to have more in common with, e.g., the expanding
envelopes of yellow supergiants (which display
HCO$^+$(3--2)/$^{13}$CO(2--1) ratios $\stackrel{<}{\sim}10$ and
CN(2--1)/$^{13}$CO(2--1) ratios $\stackrel{<}{\sim}5$; Quintana-Lacaci
et al.\ 2007) than with pre-MS disks. These results therefore are
consistent with the notion that the BP Psc disk may have its origins
not in the star formation process but, rather, in a catastrophic
interaction with a close companion during the primary's ascent of the
red giant branch (Z08). The minimum disk gas mass and angular momentum
inferred under the assumption that BP Psc is a post-MS star at a
distance of 300 pc --- i.e., $\sim$1 Jupiter mass (\S 3.2) distributed
over a disk with radius $\sim$150 AU --- is consistent with such a
companion-engulfing scenario (see discussion in \S 4.7 of Z08).

The feeble output from the BP Psc disk in the HCO$^+$ and CN lines
further implies a low molecular ionization rate, suggesting that ---
in contrast to pre-MS star-disk systems --- the BP Psc system lacks a
strong, central X-ray source. A deep {\it Chandra} observation of BP
Psc, presently scheduled for late 2008, should result in a sensitive
measurement of the ``hard'' (1.0--10 keV) X-ray flux incident on the
BP Psc disk, providing a test of this interpretation.

However, detection of a large X-ray flux from BP Psc ---
while leaving open the question of the origin of its anomalous
molecular emission line ratios --- would shed little additional light
on its evolutionary status. This is because its projected rotational
velocity may be as large as $v\sin{i} \sim 32$ km s$^{-1}$ (Z08)
which, given a radius typical of late-G giants ($\sim10 R_\odot$),
would imply a period of only $\sim2$ days.  This period is similar
both to those of rapidly rotating pre-MS stars {\it and} to those of
X-ray-luminous, G-type giants of the FK Com class (Gondoin 2005 and
references therein). The approximate upper limit on the kinematic mass
of BP Psc assuming a distance of 300 pc, 1.8 $M_\odot$ (\S 3.1), would
be consistent with FK Com status. As FK Com stars are thought to be
the products of stellar mergers (Heunemoerder et al.\ 1993 and
references therein), the comparison has interesting ramifications for
the recent history of BP Psc, under the hypothesis that it is a
post-MS star: if BP Psc is indeed a giant now engulfing a close
companion (see Soker 1998) --- forming a disk and driving jets in the
process (e.g., Nordhaus \& Blackman 2006 and references therein) ---
we may be witnessing the ``birth'' of FK Com.

\acknowledgements{The authors wish to acknowledge useful comments and
  suggestions by David Meier, David Wilner, and the 
  referee. J.H.K. thanks the staff of the Laboratoire d'Astrophysique
  de Grenoble for their support and hospitality during his yearlong
  sabbatical visit to that institution.  }


\begin{thebibliography}{}



\bibitem{} Backman, D.E., \& Paresce, F. 1993, in {\it Protostars and
    Planets III}, eds.\ E.H. Levy \& J.I. Lunine, p.\ 1253
\bibitem{} Beckwith, S., \& Sargent, A. 1993, ApJ, 402, 280 (BS93)
\bibitem{} Deguchi, S., Izumiura, H., Kaifu, N., Mao, X.,
  Nguyen-Q-Rieu, \& Ukita, N. 1990, ApJ, 351, 522
\bibitem{} Dutrey, A., Guilloteau, S., \& Guelin, M. 1997, A\&A, 317, L55
\bibitem{} Glassgold, A.E., Najita, J.R., Igea, J. 2004, ApJ, 615, 972
\bibitem{} Glassgold, A.E., Najita, J.R., Igea, J. 2007, ApJ, 656, 515
\bibitem{} Gondoin, P. 2005, A\&A, 444, 531
\bibitem{} Guedel, M. Briggs, K. R., Arzner, K., et al. 2007, A\&A,
  483, 353
\bibitem{} Huenemoerder, D. P., Ramsey, L. W., Buzasi, D. L., \&
  Nations, H. L. 1993, ApJ, 404, 316 
\bibitem{} Kastner, J.H., Zuckerman, B., Forveille, T., \& Weintraub,
  D.A. 1997, Science, 277, 67 
\bibitem{} Kastner, J.H., Vrtilek, S. D., \& Soker, N. 2001, ApJ, 550, L189
\bibitem{} Kastner, J.H., Huenemoerder, D.P., Schulz, N., Canizares,
  C.R., \& Weintraub, D.A. 2002, ApJ, 567, 434
\bibitem{} Lagrange, A.-M., Backman, D. E., \& Artymowicz, P. 2000, in
  {\it Protostars and Planets IV}, eds.\ V. Mannings, A.P. Boss, \&
  S. Russell, p.\ 639
\bibitem{} Lepp \& Dalgarno 1996, A\&A, 306, L21
\bibitem{} Nordhaus, J., \& Blackman, E. G. 2006, MNRAS, 370, 2004
\bibitem{} Omodaka, T., Kitamura, Y., \& Kawazoe, E. 1992, ApJ, 396, L87
\bibitem{} Quintana-Lacaci, G., Bujarrabal, V., Castro-Carrizo, A., \&
  Alcolea, J. 2007, A\&A, 471, 551
\bibitem{} Skatrud, D.D., De Lucia, F.C., Blake, G.A., \& Sastry,
  K.V.L.N. 1983, J. Mol.\ Spec., 99, 35
\bibitem{} Stephenson, C.B. 1986, ApJ, 300, 779
\bibitem{} Stelzer, B., \& Neuhauser, R. 2001, A\&A, 377, 538
\bibitem{} Stelzer, B., Micela, G., Hamaguchi, K., \& Schmitt,
  J. H. M. M. 2006, A\&A 457, 223
\bibitem{} Soker, N. 1998, ApJ, 496, 833
\bibitem{} Thi, W.-F., van Zadelhoff, G.-J., \& van Dishoeck,
  E.F. 2004, A\&A, 425, 955
\bibitem{} Webb, R., Zuckerman, B., Platais, I., Patience, J., White,
  R. J., Schwartz, M. J., \& McCarthy, C. 1999, ApJ, 512, L63 
\bibitem{} Zuckerman, B., Forveille, T., \& Kastner, J.H. 1995,
  Nature, 373, 494 
\bibitem{} Zuckerman, B. 2001, ARAA, 39, 549
\bibitem{} Zuckerman, B., \& Song, I. 2004, ARAA, 42, 685
\bibitem{} Zuckerman, B., et al. 2008, ApJ, in press (Z08; astro-ph/0802.0226)

\end{thebibliography}
\end{document}